\documentclass[%
 reprint,
 amsmath,amssymb,
 aps,
]{revtex4-1}

\usepackage{graphicx}
\usepackage{dcolumn}
\usepackage{bm}

\usepackage{amsmath,amssymb}
\usepackage{citesort}
\newcommand{\nn}{\nonumber}

\newcommand{\omo}{\omega_{1}}
\newcommand{\ome}{\omega_{8}}
\newcommand{\pe}{\pi_{8}}
\newcommand{\po}{\pi_{1}}
\newcommand{\ttb}{t\bar{t}}
\newcommand{\qqb}{q\bar{q}}

\begin{document}

\preprint{APS/123-QED}

\title{\boldmath Early LHC Phenomenology of Yukawa-bound Heavy $Q\bar Q$ Mesons}

\author{Tsedenbaljir Enkhbat$^{a}$, Wei-Shu Hou$^{a,b}$, and Hiroshi Yokoya$^{a,b}$}
 \affiliation{$^{a}$Department of Physics, National Taiwan University,Taipei, Taiwan 10617\\
$^{b}$National Center for Theoretical Sciences, National Taiwan University, Taipei, Taiwan 10617}


\begin{abstract}
Current limits from the LHC on fourth generation quarks are
already at the unitarity bound of 500 GeV or so.
If they exist, the strong Yukawa couplings are
turning nonperturbative, and may form bound states.
We study the domain of $m_{b'}$ and $m_{t'}$ in the range of 500 to 700 GeV,
where we expect binding energies are mainly
of Yukawa origin, with QCD subdominant.
To be consistent with electroweak precision tests,
the $t'$ and $b'$ quarks have to be nearly degenerate,
exhibiting a new ``isospin".
Comparing relativistic expansion with a relativistic bound state approach,
we find the most interesting is the production of a
color octet, isosinglet vector meson (a ``gluon-prime") via $q\bar q \to \ome$.
Leading decay modes are $\pi_8^\pm W^\mp$, $\pi_8^0Z^0$,
and constituent quark decay,
with $q\bar q$ and $t\bar t'$ and $b\bar b'$ subdominant.
The color octet, isovector pseudoscalar $\pi_8$ meson decays
via constituent quark decay, or to $Wg$.
These decay rates are parameterized by
the decay constant, the binding energy and mass differences, and $V_{tb'}$.
For small $V_{t'b}$ , one could have a spectacular signal of
$WWg$, where a soft $W$ accompanies a very massive $Wg$ pair.
In general, however, one has high multiplicity signals
with $b$, $W$ and $t$ jet substructures that are
not so different from $t'\bar t'$ and $b'\bar b'$ search.
\begin{description}
\item[PACS numbers]
14.65.Jk 
11.10.St 
13.85.Rm 
13.25.Jx 
\end{description}
\end{abstract}

\pacs{Valid PACS appear here}
\maketitle


\section{\label{sec:Intro}INTRODUCTION\protect\\}

The CMS experiment at the Large Hadron Collider (LHC)
announced~\cite{Tonelli} recently that,
for Standard Model (SM) with four quark generations (4G),
``the Higgs boson in the mass range of 120 to 600 GeV is
excluded at the 95\% C.L.".
The ATLAS experiment is in agreement~\cite{Nisati}.
A common 
inference is that the
4G itself is 
practically ruled out. 
Afterall, CMS also reported the most stringent bounds on
the $t'$ and $b'$ quarks to date:
$m_{t'} > 450$ GeV~\cite{Tonelli} and $m_{b'} > 495$ GeV~\cite{DeRoeck},
both at 95\% C.L., which are rather close to
the unitarity bound (UB) of 500--550 GeV~\cite{Chanowitz:1978uj}.
However, in as much as 4G may not exist,
an intriguing possibility~\cite{Bardeen:1989ds, Holdom:1986rn, Holdom:2009rf} is
that electroweak symmetry breaking (EWSB) itself might be
triggered by, or related to, the strong Yukawa couplings of
the $t'$ and $b'$ quarks.
Can the UB violation of strong $WW$ scattering~\cite{Lee:1977yc}
be related to the UB violation of strong $Q\bar Q$ and $QQ$ scattering?
To pursue such questions is a major purpose of the LHC,
and is well within its abilities.

If a relatively light Higgs particle emerges soon at the LHC
with SM cross sections, then 4G would truly be in trouble~\cite{Djouadi}.
But, the exclusion statement of SM/4G Higgs might well
get extended to the SM Higgs itself with 2011-2012 LHC data.
For Higgs particle beyond 600 GeV or so, one enters the
strong $WW$ scattering domain, and the ``Higgs" becomes
a broad object~\cite{Lee:1977yc, PDG}, which requires both
high energy (14 TeV) and high luminosity to explore.
For such a heavy Higgs boson, if the $t'$ or $b'$ quarks were
however found below 500--550 GeV or so,
then the Yukawa sector may not be strongly coupled enough
to link with the strongly coupled ``Higgs sector".
Thus, we have in mind the scenario where
neither the (SM-like) Higgs boson, nor the 4G quarks,
are found below 600 GeV and 500--550 GeV, respectively.
%

%
%
It is important to remember that new $CP$ violating phases
associated with 4G quarks may link to~\cite{Hou:2008xd}
the Baryon Asymmetry of the Universe (BAU).
Thus, the existence of a very heavy 4G may touch both
EWSB and BAU, which are two of the greatest problems in particle physics.
This provides strong motivation for continuing the pursuit of
4G in this volatile time.

If the Higgs boson is heavy (and ``fat"), while the 4G quarks
are above the UB, then whether the large Yukawa coupling
induces $Q\bar Q$ condensation~\cite{Holdom:2009rf} or not, it would be
important to explore possible \emph{bound states} of this strong coupling.
This is not just about potentially interesting LHC phenomenology,
but may be necessary to provide a guide for
the search of ultraheavy chiral quarks beyond UB.
The main purpose of this paper is to explore
lower lying bound states of strong Yukawa coupling,
and the associated properties.
However, by venturing above the UB,
one immediately encounters the perils of the
breaking down of perturbation theory. Thus,
in lieu of genuine nonperturbative approaches,
such as~\cite{dlin} lattice field theory (LFT), our work is only of an
illustrative kind.

One aid to the study is a new, heavy isospin.
If 4G quarks exist, by which we mean a sequential
left-handed doublet and a pair of right-handed singlets
under the weak interaction, the $S$, $T$ variables~\cite{Peskin:1990zt}
or electroweak precision tests (EWPrT) require
the $t'$ and $b'$ to be nearly degenerate (and here,
independently, a heavier Higgs is also required~\cite{Kribs:2007nz}).
Of course, some small splitting is needed to satisfy $T$,
but in this paper we will treat the $t'$ and $b'$ as degenerate,
hence one has a new ``isospin".
This isospin, in contrast to the chiral limit of $u$ and $d$
quarks under QCD, is in the opposite limit of
degenerate \emph{ultraheavy} quarks.
We thus borrow the notation of isovector $\pi$, $\rho$
(or $[\bar t'b',\ (\bar t't'-\bar b'b')/\sqrt{2},\ \bar b't']$)
and isoscalar $\eta$, $\omega$ (or $(\bar t't'+\bar b'b')/\sqrt{2}$)
etc., for the heavy $\bar QQ$ ``mesons".
As we will see, unlike technicolor, the ``$\rho$" meson
does not play a major role for Yukawa bound states, nor does the $\eta$.
Another interesting point is that,
if the Yukawa interaction is
the dominant binding mechanism,
since it is color blind, the $\bar QQ$ mesons,
unlike the QCD situation, come not only in color
singlets, but color octets as well~\cite{Wise}.
We find the states $\omega_8$, $\omega_1$, $\pi_8$ and $\pi_1$,
where the subscript indicates the color representation,
to be of phenomenological interest.
In particular, $\omega_8$ and $\pi_8$ may be
accessible in the near future.

In Sec. II, we bring forth the issues of strongly coupled
relativistic bound states. We contrast the
necessarily relativistic $\gamma_5$ coupling of
weak Nambu--Goldstone boson (NG),
or longitudinal vector boson exchange,
to the Coulombic QCD bound state, as well as scalar exchange.
Because the Higgs scalar should be treated as heavy now,
its effect is less prominent than NG exchange.
We compare the traditional relativistic expansion~\cite{Wise}
with a relativistic Bethe-Salpeter approach~\cite{Jain},
and illustrate why the Yukawa bound state involves
highly relativistic motion of its constituents.
An issue appears regarding the treatment of
$s$-channel NG exchange, towards strong Yukawa binding.
This, plus other issues, forces us to compromise in the
study of the possible spectrum at this stage,
and we restrict ourselves to $m_Q$ in the range of
500--700 GeV, i.e. not far above the UB. In this way,
the relativistic expansion provides a partial guide,
while we offer a peek beyond and consider binding energies
of order $-100$ GeV or more, for the possible spectrum
around and above the TeV scale.
This leads us to focus on, from the production point of view,
an $\omega_8$ state (effectively a ``$g'$", or gluon-prime),
as well as possibly the $\pi_8$ and $\omega_1$ states in its
decay final state.

In Sec. III we turn to exploring the production and
decay properties. We survey the key parameters needed:
the binding energy, the vector--pseudoscalar and octet--singlet mass splittings,
the vector meson decay constant $f_{\omega_8}$, and
the quark mixing element $V_{t'b}$.
Production is mostly through $q\bar q \to \omega_8$
and depend only on $f_{\omega_8}$, but
the annihilation, transition, and free quark decays
involve all these parameters, where
the numerical values we use are only illustrative.
We find, in general, that the $\omega_8$ is relatively narrow,
but has a host of decay final states, which might
therefore elude early detection.
We offer some discussion of the phenomenology at LHC in Sec.~IV,
touching briefly on \emph{deep} bound states, i.e.
the possibility of binding energy approaching $m_Q$ itself,
for $m_Q$ beyond 700 GeV.
We end with a conclusion and outlook.

\section{\label{sec:II}Strongly Coupled Relativistic Bound States\protect\\}

When the unitarity bound is reached for very heavy chiral quarks,
it means that some \emph{tree level} $Q\bar Q$ and $QQ$ scattering
cross sections will violate unitarity, or conservation of probability,
in the high energy limit.
Pointed out over 30 years ago~\cite{Chanowitz:1978uj},
it is remarkable that we are now at the doorsteps
to probe whether such new heavy chiral quarks exist.

Many, if not most people, tacitly take the UB as a kind of
ceiling for SM-like chiral quarks to make sense.
But in reality, crossing the UB just implies that
the Yukawa couplings are becoming so strong,
they are turning nonperturbative.
We have seen how the remarkable theory of QCD turns nonperturbative
in the infrared, resulting in the rich phenomena of hadrons.
We will not dwell on deeper short-distance (UV) gauge theories like
technicolor, but just take the large Yukawa couplings~\cite{upper} at face value:
if chiral quarks $Q$ (a left-handed doublet of $t'$ and $b'$ in our case)
exist at or above the UB, what could be the emergent phenomena?

Consider first heavy quarkonia bound by QCD.
Since QCD is perturbative at short distance,
we have the familiar Coulombic bound states with
a $1/\beta$ enhancement factor, where $\beta = v$ is the velocity.
This is the domain of Non-Relativistic QCD (NRQCD), where one
expands in velocity, which is of order $\alpha_S$.
The NR nature makes good contact with the familiar atomic systems.

Exchanging Higgs bosons brings in the Yukawa couplings,
which has been considered in the literature.
For our case, we will treat the Higgs boson as
above~\cite{Tonelli, Nisati} 600 GeV and heavy,
which suppresses the binding effect due to Higgs exchange.
However, NG or longitudinal vector boson exchange
(transverse gauge boson exchange has coupling constant
$g$ or $e$, hence subdominant and largely ignored by us)
also couples with the Yukawa coupling strength, but it
involves the $\gamma_5$, which couples the upper and lower
components of the massive Dirac quark. Since the lower
component vanishes when the heavy quark is at rest,
NG exchange is suppressed in the NR limit.
Conversely, the coupling to high momentum heavy quarks
is large, the more so the heavier the quark.
This reflects the derivative coupling nature of
longitudinal vector bosons.
The upshot is that, Yukawa interactions between very
heavy quarks are large when these quarks are in
relativistic relative motion, i.e. with momentum $q \sim m_Q$.

With this insight, and ignoring the Higgs exchange for
the moment, if the bound state formation is dominated
by QCD, then the NR nature of the bound state
($\beta \simeq \alpha_S$) actually suppresses the
effect of Yukawa coupling.
However, as the Yukawa coupling increases with $m_Q$,
although the QCD-bound system becomes even more nonrelativistic,
at some point NG exchange would (perhaps suddenly) take over,
and one would find the bound state system turns ultrarelativistic.

We shall illustrate with two different perspectives,
one a traditional relativistic expansion~\cite{Wise}, the other
a relativistic Bethe-Salpeter approach~\cite{Jain}.

\subsection{Relativistic Expansion}

A standard approach in considering bound state phenomena
is to make a relativistic expansion around the leading
potential.
The Higgs potential for very heavy quarks, in the context
of forming bound states, was considered a long time ago~\cite{Inazawa:1987vk}.
The relativistic corrections were recently calculated in Ref.~\cite{Wise}.
The scattering amplitudes for $t$-channel Higgs, neutral and charged
NG (called fictitious scalar in Ref.~\cite{Wise}), and gluon exchange,
as well as $s$-channel NG exchange, were computed.
Ref.~\cite{Wise}, however, did not put in
$s$-channel Higgs and gluon exchange, even though both
color singlet and octet $Q\bar Q$ configurations were discussed.
Touching both $m_H = 130$ GeV and $m_H = m_Q$ cases,
a variational approach was used to estimate the size $a_0$
(equivalent to wavefunction) and binding energy for
$I = 0,\ 1$, $S = 0,\ 1$, color singlet and octet states.

We will not repeat what was already done here,
but just give some salient features.
As a control on validity of the relativistic ($v$ or $\beta$)
expansion, the authors of Ref.~\cite{Wise} required
$\textbf{q}^2/m_Q^2 < 1/3$ ($|\textbf{q}| \equiv q$ is the
relative momentum), which translates into $a_0 > \sqrt{3}/m_Q$.
Since $t$-channel gluon exchange gives the Coulomb potential,
it is clear that one has a Coulombic QCD bound state
when $m_Q$ is not yet too large,
with Bohr radius $a_{\rm QCD}$. 

The low mass Higgs case is no longer tenable with 4G,
both by direct search~\cite{Tonelli, Nisati},
and indirectly~\cite{Kribs:2007nz} from EWPrT due to the heaviness of 4G quarks.
For the heavy Higgs case illustrated with $m_H = m_Q$,
 as one can see from Fig.~8 of Ref.~\cite{Wise},
the radii $a_0$ of the $\omega_1$, $\rho_1$ and $\eta_1$
(the subscript $1$ stands for color singlet) states
decrease rather slowly below $a_{\rm QCD}$ as $m_Q$ increases.
The $\pi_1$ radius rises slowly above $a_{\rm QCD}$ due to
the extra repulsion it receives from $s$-channel NG exchange.
The Yukawa effect is subdominant compared with QCD, where
inspection of the binding energies offer further support:
they rise slowly from $\sim -1$ GeV around $m_Q = 200$ GeV,
to $\sim -2$ GeV at $m_Q = 300$ GeV, with $\pi_1$ only
slightly lower.
However, just before $m_Q$ reaches 400 GeV,
radius $a_0$ for $\omega_1$ drops precipitously, while
the binding energy rises sharply from around $-2.5$ GeV,
to around $-100$ GeV at $m_Q = 500$ GeV.
The condition $a_0 > \sqrt{3}/m_Q$ is violated shortly above $m_Q = 400$ GeV.

The ``kink" around 400 GeV is the point where the
NG exchange has wrestled the mechanism for binding
away from the usual NR QCD binding.
The sudden drop of the size of the bound state is
due to tapping the large attraction at large momentum
$q$ for the bound quark (besides the $\gamma_5$ coupling
of the NG bosons, the heavy Higgs also defines a very
short range for the strong Yukawa coupling to be effective).
The resulting larger binding energy overcomes
the much increased kinetic energy.
For $\rho_1$, the onset is delayed until $m_Q = 530$ GeV or so,
with binding energy of $-25$ GeV at $m_Q = 600$ GeV.

For color octet states, which are not bound by QCD,
the binding energy for $\omega_8$, due purely to Yukawa coupling,
turns on sharply around $m_Q = 535$ GeV,
rising to $-35$ GeV for $m_Q = 600$ GeV.
The $\rho_8$ state turns on much later, around $m_Q = 700$ GeV.
But, unlike $\pi_1$, because there is no $s$-channel repulsion,
$\pi_8$ is degenerate with $\omega_8$.
It should be noted that the Yukawa effects of neutral and charged
NG exchange is weaker but constructive for $\omega_8$,
while the converse is true for $\pi_8$,
so this degeneracy could be accidental.
Furthermore, this degeneracy should be lifted by $s$-channel
gluon annihilation, which would raise $m_{\ome}$ but
was not considered by the authors of Ref.~\cite{Wise}.
As we shall soon see, the vector channel also has
$S$- and $D$-wave mixing.

Although identifying the $\omega_1$ as the
lightest color singlet, and $\omega_8$ (and $\pi_8$) as the
lightest color octet, it is ironic that the
relativistic expansion breaks down almost immediately after
the strong Yukawa binding becomes potent.
But it does illustrate that one needs a genuine relativistic
approach in treating strong Yukawa binding.
We turn to such an approach that is in principle
nonperturbative, but carrying its own dubious features:
the Bethe-Salpeter (BS) equation.

\subsection{Relativistic Bethe-Salpeter Approach}

A long time ago, around the time the SSC (Superconducting
Super Collider) was under construction but then unfortunately
canceled, the authors of Ref.~\cite{Jain} pursued the
BS equation approach to the relativistic bound states of
very heavy sequential 4th generation quarks.
It consists of a ladder sum of the scattering amplitudes
that appear in the relativistic expansion.
In the heavy isospin limit and treating $M_Z = M_W$,
a clear isospin reorganization separates
into $I = 0$ and $I = 1$ ``mesons".

For $\bar QQ$ meson with total  momentum $P$ and
relative momentum $q$, one has a set of integral equations,
with loop momentum $q'$, where $q'-q$ is the momentum exchange
in $t$-channel. However, for $s$-channel annihilation contribution,
the annihilation momentum is $P$ itself,
and the integral over loop momentum $q'$ carries
no $q$ dependence, giving a possibly divergent constant.
To remedy this, Ref.~\cite{Jain} took a fixed $q$ subtraction at $q = 0$.
In this way, all the $s$-channel diagram contributions
were eliminated from the BS equation.
This includes even the $s$-channel gluon exchange
for the octet, isosinglet vector channel,
which was not considered in Ref.~\cite{Jain}, as
the authors concerned themselves with color singlet states only.

In terms of mathematical physics, to set up
integral equations to be solved in a self-consistent way,
the subtraction at constant $q$ seems reasonable.
However, as admitted by the authors of Ref.~\cite{Jain}
in a footnote, the $s$-channel NG exchange leads to
repulsion. Thus, in discussing bound state solutions,
there is the issue of the physical correspondence,
and therefore the range of validity (in $m_Q$)
for implementing the subtraction.
In the relativistic expansion, one clearly would not
drop the $s$-channel diagrams.

Our purpose here is not to make a full treatment of the BS equation,
as it is only a ladder sum of  $t$-channel exchange diagrams,
with higher order corrections ignored.
Furthermore, while the BS equation is relativistic,
its solution depends very much on the approximations and ansatz made.
Ref.~\cite{Jain} illustrated with covariant gauge
(but employing a weak coupling relation between
spatial and temporal spinor components), the
instantaneous approximation with positive frequency potential only,
as well as keeping both positive and negative frequency potentials.
Although the numerical solutions share common features,
the bound state mass ($M$) values differ.
Starting all from $M \cong 2m_Q$ for low $m_Q$,
they decrease smoothly as $m_Q$ increases,
without exhibiting the kink seen in the relativistic expansion.
As such, the BS approach is an improvement.
But, as a common feature, once the binding energy
$M - 2m_Q$ becomes significant (e.g. $-100$ GeV or so),
at high enough $m_Q$, the low lying mesons collapse.
That is, the binding energy becomes so large
such that the total mass drops to zero.

\begin{figure}[t!]
\centering
{ \includegraphics[width=80mm]{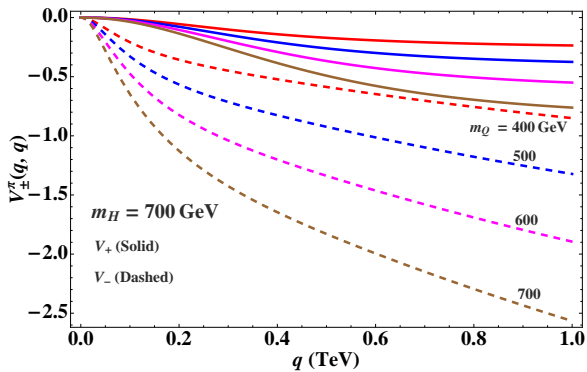}
}
\caption{
The $V^\pi_+$ (solid) and $V^\pi_-$ (dashed) potentials
in Eq.~(1) for the $\pi_1$ state,
for $m_Q = 400,\ 500,\ 600,\ 700$ GeV, and $m_H = 700$ GeV,
plotted along $q' = q$.
Both vanish for $q = 0$, or zero relative motion,
but go to rather large values for large $q$,
with $V^\pi_-$ turning on rather sharply for
$q \gtrsim 100$ GeV.
} \label{Vpm}
\end{figure}

Having bound state solutions numerically turning tachyonic
for a strongly coupled system is not particularly astounding.
For QCD (and likewise for QED),
quarkonium masses calculated at fixed order
could also vanish at large enough coupling strength.
The system has turned fully relativistic with strong coupling,
and the familiar Bohr-Schr\"odinger solution is no longer valid.
For the BS equation, however, it is already relativistic.
The collapse of the meson state may be related to
the symmetry breaking itself~\cite{Kugo}, but
because of the approximate nature of the BS equation,
as well as the numerical approximations made in its solution,
we refrain from dwelling further on this.

%

Rather, we wish to use the BS equation and its numerical study
to compare and contrast with the previously discussed relativistic expansion,
to project what may really happen for relativistic, strong Yukawa bound states,
before any ``collapse", such as illustrated in Ref.~\cite{Jain}, could occur.
For this purpose, we note that with the
\emph{subtracted} BS equation, hence with the
$s$-channel repulsion removed,
the $\pi_1$ turns out to be the most attractive system,
i.e. the lightest bound state:
the NG, gluon and Higgs exchange are all attractive,
as can be seen also from Ref.~\cite{Wise}.
The behavior for $\omega_1$ is indeed similar, but
the formalism is more complicated for the $1^-$ system,
where the $S$- and $D$-wave channels are coupled.
Note that Ref.~\cite{Jain} used (sometimes tacitly)
a Higgs mass around 100 GeV, which is no longer valid.
In our mind, we are interested in the bound state dominated by
Yukawa coupling, i.e. by NG exchange, hence we will
view gluonic exchange as correction, with Higgs exchange
perhaps even milder. This matches to what one finds in
the relativistic expansion, that the Yukawa binding
suddenly turns on, bringing on a rapid rise in
binding energy.
In effect, we use the BS equation approach to check, and probe beyond,
the ``kink" seen in the relativistic expansion.

The equation for $\pi_1$ is the most compact.
In the impulse approximation, integrating over $q_0$
gives the amplitude $\chi(q)$, where $q = |\textbf{q}|$
is the relative momentum.
Keeping both positive and negative frequency
amplitudes $\chi_\pm(q)$, one has the coupled equations,
\begin{eqnarray}
(M \mp 2\omega)\chi^{\pi}_\pm(q) &=&
 \pm \int dq^\prime \frac{q^\prime}{q} \left[V^{\pi}_\pm(q,\,q')\chi^{\pi}_+(q^\prime) \right.
  \nonumber \\
 && \hskip1.4cm \left. +\, V^{\pi}_\mp(q,\,q')\chi^{\pi}_-(q^\prime)\right],
\label{BSpi1}
\end{eqnarray}
where $M$ is the eigenvalue,
and $\omega = \sqrt{m^2 + q^{2}}$.
The potentials $V^{\pi}_\pm(q,\,q')$, where we have absorbed a
factor of $1/\pi$ into its definition as compared to Ref.~\cite{Jain},
arise from $t$-channel diagrams as described earlier.

Let us understand the $V_+$ and $V_-$ potentials.
The less familiar one is $V_-$, which
couples $\chi_\mp$ to $\chi_\pm$, while $V_+$ is ``diagonal".
In the limit that $|V_+| \gg |V_-|$, one simply has
$M = 2\omega + \langle {V_+}\rangle$, where $\langle {V_+}\rangle$ is
analogous to the expectation value of the potential ($\chi_+(q)$ is
like $\psi^\dag\psi$) of nonrelativistic quantum mechanics.
On the other hand, if $2\omega \gg |V_-| \gg |V_+|$,
then $M \simeq 2\omega - V_-^2/4\omega$,
hence the $V_-$ contribution is more suppressed than
the corresponding $V_+$ contribution when it is weak.

The potentials $V_\pm(q,\,q')$ are symmetric in $q$ and $q'$,
and is steepest along $q' = q$.
We plot $V^\pi_\pm(q,\,q)$ in Fig.~\ref{Vpm},
for heavy $m_H = 700$ GeV, and for $m_Q = 400$, 500, 600, 700 GeV.
We have checked that the $V^\pi_+$ potential drops by
a factor of two or more as $m_H$ moves from 100 to 700 GeV,
but $V^\pi_-$ is rather insensitive to $m_H$.
We see from Fig.~\ref{Vpm} that both the $V^\pi_+$ and $V^\pi_-$
potentials are suppressed for low relative momentum $q$,
agreeing with the relativistic expansion view.
As $q$ increases, $V^\pi_+$ increases relatively slowly,
reaching $-V^\pi_+ \simeq 0.3\; (0.7)$ at $q = 500$ (700) GeV
for $m_Q = 500$ (700) GeV.
But $V^\pi_-$ turns on more sharply,
reaching beyond $-V^\pi_- \simeq 1$ at $q = m_Q = 500$ GeV,
and $-V^\pi_- \simeq 2$ at $q = m_Q = 700$ GeV.

So what does the strength of $V^\pi_\mp$ mean?
Take the $\chi^\pi_+$ equation,
the binding energy $E = M - 2m_Q$ is equal to $2\omega - 2m_Q$,
the kinetic energy due to motion,
plus the right hand side of Eq.~(\ref{BSpi1}),
which should be the ``potential".
Let us normalize $q'$ by $m_Q$.
Since $V^\pi_\pm(q,\,q')$ is steepest for $q' = q$,
we see that $m_Q|V^\pi_\mp(q,\,q)| \simeq m_Q$ for $q\simeq m_Q$
means the ``potential" is comparable to $m_Q$ in strength
when the momentum is comparable to the rest mass,
and the kinetic energy due to motion is $2(\sqrt{2} -1)m_Q$.
We can now sense why there is a tendency for the $\pi_1$ state
to collapse already for $m_Q$ not far above 500 GeV
in the numerics of Ref.~\cite{Jain}.

\subsection{\boldmath Possible Spectrum for $m_Q \simeq 500$--700 GeV}

We have illustrated how the binding energy for the (heavy) isovector,
color singlet $0^-$ state, which we call $\pi_1$, could be already
moving towards collapse for $m_Q$ as low as 500 GeV
(this number from Ref.~\cite{Jain} contains the attraction
 due to light Higgs exchange),
if the BS approach of Eq.~(\ref{BSpi1}) holds true.
However, this state receives a repulsive $s$-channel NG
annihilation contribution. Furthermore, as the tendency
towards collapse approaches, the ladder sum BS equation
may no longer be sufficient. Even within the BS approach,
where Eq.~(\ref{BSpi1}) gives rise to the earliest collapse,
if one drops the negative frequency amplitude, the collapse
is delayed~\cite{Jain} by almost a factor of two in $m_Q$.
In the covariant gauge but using a weak coupling relation
between temporal and spatial spinor components, the collapse of $\pi_1$
occurs slightly before the positive frequency only case.
It is not clear at what $m_Q$ the collapse
truly occurs numerically. In any case,
we would not touch the collapse here,
as it is not yet well understood.
It seems prudent, then, to consider binding energies of
$-100$ to $-200$ GeV, but not more,
to make a preliminary study of possible phenomena at LHC,
running at $\sqrt{s} = 7$ TeV.

A similar equation as Eq.~(\ref{BSpi1}) holds for the
isosinglet, color singlet $0^-$ state, the $\eta_1$.
Although gluon and Higgs exchange are attractive,
NG exchange turns repulsive~\cite{Jain, Wise}.
Thus, $\eta_1$ is much heavier than $\pi_1$,
and cannot be a low-lying state.

Turning to vector mesons, Ref.~\cite{Jain} found,
similar to the relativistic expansion of Ref.~\cite{Wise},
that the isoscalar $\omega_1$ has the tightest binding,
though it is weaker than ($s$-channel subtracted) $\pi_1$.
This concurs with our earlier observation that
Yukawa effects are weaker but constructive for vector,
while stronger but destructive for pseudoscalar.
However, there is no repulsive $s$-channel effect for $\omega_1$.
On the other hand, checking the formalism,
we find that there is $^3S_1$-$^3D_1$ mixing,
resulting in a set of more complicated coupled BS equations.
From the numerics of Ref.~\cite{Jain}, we expect that
for $m_Q$ in the range of 500--700 GeV, binding energies
of order $-100$ to $-200$ GeV is reasonable for $\omega_1$.
The $\rho_1$ state, analogous to $\eta_1$ in receiving
the repulsive NG exchange, is far less binding,
which is also seen from the relativistic expansion.

Turning to color octet counterparts, one can treat gluon exchange
as a perturbation when binding energies are much large than
QCD bound states. Given that the bound state should be rather small,
the repulsion due to gluon exchange should be larger than typical QCD binding.
But without fully solving for the bound state,
octet-singlet splitting is uncertain.
However, there should be no doubt that $\pi_8$ and $\omega_8$
would be the low-lying octet states, in agreement with the
relativistic expansion. The octet $\eta_8$ and $\rho_8$
are likely not (or at best weakly) bound,
hence we do not consider these.

To summarize, the states to keep in mind are
$\pi_1$, $\omega_1$, $\pi_8$ and $\omega_8$.
There are possibly other states with different quantum numbers,
but in general they would not be lighter, while
likely possessing more complicated properties.
The absence of $\rho$ states is distinct from
QCD-like gauge theories such as technicolor.
For $m_Q$ in the range of 500 to 700 GeV,
staying short of very tight binding,
these states would probably populate the 1 to 1.4 TeV range,
with binding energies of order $-100$ to $-200$ GeV.
The ordering of the spectra, according to Ref.~\cite{Jain}
(which did not actually consider octet states),
would be $M_{\pi_1} \lesssim M_{\omega_1}
 \lesssim M_{\pi_8} \lesssim M_{\omega_8}$.

Since we are concerned with LHC phenomenology
and the heavy quark search program in the near future,
we should consider briefly issues regarding production:
\begin{itemize}
 \item $\pi_1$ and $\omega_1$ cannot be produced via $gg$ fusion,
   but can be produced via weak Drell-Yan processes,
   hence have a weak production cross section;
 \item $\pi_8$ can be pair produced by $gg$ and $q\bar q$
   scattering~\cite{Dobrescu:2011px}, but it is heavier
   and less efficient at 7 TeV.
\end{itemize}
This leaves $\omega_8$, which has the same quantum numbers as
the gluon, 
that is the most attractive in the
near future in terms of production cross section.
It cannot be produced by $gg$ fusion, as two massless vectors
cannot forge a massive vector (Yang's theorem) particle,
hence the production is limited to $q\bar q$ fusion.
In the next section, we turn to the production
and decay properties of the $\omega_8$ meson.
Note that if an $\eta_8$ or $\eta_1$ state existed,
such as for QCD binding,
it could tap $gg$ production~\cite{Arik:2002nd}.

\section{\label{sec:omega8} \boldmath
 Production and Decay of $\omega_8$ \protect\\}

In this section, we discuss the production and decay of the $\ome$
meson, which we have portrayed as the likely leading harbinger of 4G
bound-states in our scenario, where a very heavy quark
{\it isospin} symmetry prevents the production of
some mesons directly at hadron colliders.

For the numerical study of the production cross-sections and decay
rates, the following parameters are needed as input:
the decay constant $f_{\ome}$,
the 4G quark masses $m_Q = m_{t'} = m_{b'}$ (we assume
 degeneracy of $t'$ and $b'$ as a simplifying approximation,
 hence an isospin symmetry),
the mass of resonances, {\it i.e.} the binding energy of resonances,
and finally the quark mixing elements $|V_{t'b}| \cong |V_{tb'}|$.
Once the constituents as well as their interactions are fixed,
the decay constant and binding energy are the consequence of
the dynamics of the system.
As we stressed previously, in the range of
strong Yukawa couplings we consider, their estimation has to be
done by LFT approach~\cite{dlin} for more quantitative understanding.
However, such an endeavor is beyond the scope of this paper.
Instead we parameterize them in our phenomenological study.

The decay constant of $\ome$ is defined through
\begin{align}
 \langle 0|V^{\mu,a}|\ome^b(p)\rangle \equiv
 \frac{1}{\sqrt{2}}\delta^{ab} f_{\ome} m_{\ome}
 \varepsilon^{\mu}(p),
 \label{eq:dec}
\end{align}
which we parameterize by a dimensionless parameter
$\xi = f_{\ome}/m_{\ome}$.
The mass $m_{\ome}$ is the most important,
as we discuss the production and decay of $\ome$.
We also assume $|V_{t'b}| = |V_{tb'}|$ is the dominant quark
mixing element, and ignore mixings with lighter generations.

\begin{figure}[ht]
\begin{center}
 \includegraphics[width=30mm]{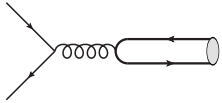}
\caption{Production mechanism for the color octet, isoscalar-vector
 meson $\ome$ at hadron colliders.}\label{fig:prod}
\end{center}
\end{figure}

\subsection{Production}\label{sub:prod}

The dominant production mechanism for $\ome$ is,
as illustrated in Fig.~\ref{fig:prod}, via $\qqb$ annihilation.
Because one cannot fuse two massless gluons into a
massive vector boson, gluon-gluon fusion is not operative.
We have also computed the higher order $gg\to\ome g$ process
as a check.

At the parton level, the total cross section is
\begin{align}
 &\hat\sigma_{\qqb\to\ome}(\hat{s}) =
 \frac{32\pi^3\alpha_s^2}{9m_{\ome}^2}\xi^2
 \delta\left(1-m_{\ome}^2/\hat{s}\right). 
\end{align}
%
%
%
Convolving with the parton luminosity
\begin{align}
 {\cal L}(\tau;\mu_F^2) = \int\int dx_1 dx_2 f_1(x_1;\mu_F^2)
 f_2(x_2;\mu_F^2) \delta\left(\tau-x_1x_2\right),
 \nonumber
\end{align}
we get the hadronic cross section,
\begin{align}
 \sigma(s)=\int d\hat\tau\,\hat\sigma(\hat\tau s)\,{\cal
 L}(\hat\tau;\mu_F^2).
\end{align}

\begin{figure}[t]
 \begin{center}
 \includegraphics[width=81mm]{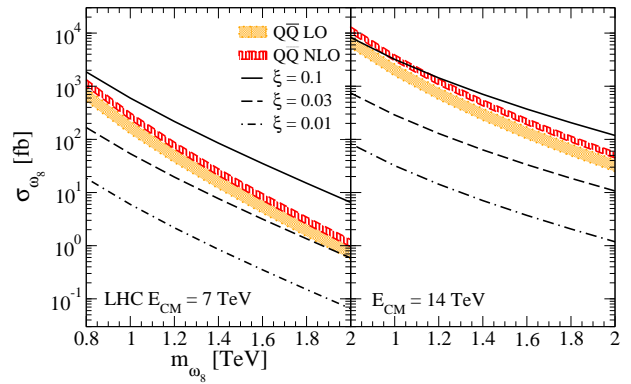}
 \caption{Production cross-section of $\ome$ at the LHC
 running at 7~TeV (left) and 14~TeV (right)
 for various $\xi = f_{\ome}/m_{\ome}$ values.
 The open $Q\bar Q$ cross sections at LO and NLO
 are also plotted for comparison.}
  \label{fig:pro}
 \end{center}
\end{figure}

We plot on the left-hand side of Fig.~\ref{fig:pro} the cross section
for inclusive $\ome$ production at the LHC with $\sqrt{s}=7$~TeV.
We use CTEQ6L~\cite{Pumplin:2002vw} parton distribution functions (PDFs), and set
the renormalization and factorization scales to $\mu_R=\mu_F=m_{\ome}$.
We use three values of $\xi = f_{\ome}/m_{\ome} = 0.1$, 0.03 and 0.01
for illustration.
Although we do not have any suggestive estimation for the decay constant,
since the Yukawa bound state is highly relativistic,
we expect the larger (smaller) value of
$\xi$ corresponds to a stronger (weaker) bound meson.
We note that the decay constant divided by meson mass
for usual $\rho$, $J/\psi$ and $\Upsilon$ are ${\cal O}(0.1)$.
However, we are not dealing with usual QCD-bound
meson production~\cite{Baier:1983va, Barger:1987xg},
so we leave $\xi$ as a parameter.
The cross section is proportional to $\xi^2$ and
decreases with increasing $m_{\ome}$.
Because of our ignorance of the decay constant $f_{\ome}$,
the cross section $\sigma_{\ome}$ ranges from pb to fb,
for $m_{\ome}$ ranging from 900 to 1400 TeV.
The plot extends to 2 TeV, since it depends only on $\xi$,
which we view should be experimentally determined.
As the gluon density at the LHC is large,
we have checked the higher order $gg\to\ome g$ scattering process,
and find the contributions to be quite small
for the region of our interest, $m_{\ome}>0.8$~TeV.

For comparison, we also plot the open $Q\bar Q$
pair production cross section at LO and NLO~\cite{Nason:1987xz},
as a function of $2m_Q$, matching (approximately)
to $m_{\ome} = 2m_Q$ on the same plot.
The cross section should be multiplied by two to take into
account the production of the degenerate 4G doublet.
In NLO calculation, we use CTEQ6M PDFs~\cite{Pumplin:2002vw}.
To see the uncertainty of the cross section,
as well as the size of NLO correction, we vary the
scales in each calculation from $\mu_R=\mu_F=m_Q$ to $4m_Q$.
The uncertainty for LO (NLO) prediction is expressed as
the dotted (hatched) band.

At the LHC running at 7~TeV, the $\ome$ production
cross section with $\xi=0.1$ is about the same
order as twice the open $Q\bar{Q}$ production cross section in
the lower mass region, but exceeds the latter in the higher mass region.
On the other hand, when $\xi$ is smaller, the cross section
could be well below the open production cross section.
Thus, $\ome$ production can be interesting
if $\xi$ is sizable, such as of $0.1$ order.
Note that $\omega_8$ is produced through $q\bar q$, while
open $Q\bar Q$ is produced dominantly through $gg$ fusion.

We plot on the right-hand side of Fig.~\ref{fig:pro}
the $\ome$ cross section for LHC at $\sqrt{s}=14$~TeV,
which is an order of magnitude larger than those for LHC at 7~TeV.
The open production cross section grows relatively larger than the
$\ome$ production because of increase in gluon luminosities.

\subsection{Decay}\label{sub:dec}

The decay channels of $\ome$ we consider are
\begin{itemize}
 \item Annihilation Decay:
       $\ome\to\qqb$, $\ttb$; $t\bar{t}'$, $b\bar{b}'$;
 \item Free-quark Decay:
       $\ome\to bW\bar{t}'$, $tW\bar{b}'$;
 \item Meson Transition:
       $\ome\to \omo g$; $\pe W$.
\end{itemize}
In the following, we discuss each decay mode separately.

\subsubsection{Annihilation Decay}\label{sub:ann}

\begin{figure}[t!]
\begin{center}
 \includegraphics[width=30mm]{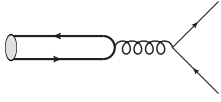}
 \caption{QCD-induced annihilation decay of $\ome$ 
  into light quark pair (dijets) or $\ttb$. 
  }
 \label{fig:anni}
\end{center}
\end{figure}

First, $\ome$ can decay into dijets or $\ttb$ by the
co-annihilation of the 4G quarks inside the bound state
through QCD, as shown in Fig.~\ref{fig:anni}.
These are the reverse processes of the production mechanism,
therefore the existence of these decay modes is robust.

The decay partial width in this mode is proportional to $\xi^2$.
The two-body decay width is calculated at Born level to be
\begin{align}
 \Gamma(\ome\to\qqb) & = \xi^2 \frac{\pi\alpha_s^2}{3}\, m_{\ome} n_f, \\
 \Gamma(\ome\to\ttb) & = \xi^2\frac{\pi\alpha_s^2}{3}\, m_{\ome} {\beta}_t,
\end{align}
where $n_f=5$ is the number of light quark flavors,
and ${\beta}_t = \sqrt{1-4m_t^2/m_{\ome}^2}$ is
the velocity of the top quark in the $\ome$ rest-frame.
Due to the number of light quark flavors,
the decay partial width into dijets is
larger than that into $\ttb$ by $\sim 5$.

Analogous to $gg\to \ome g$ production, we have also
estimated the three-body $\ome \to ggg$ 
decay rate, following the tree-level calculation of
Ref.~\cite{Petrelli:1997ge}, and found that it can be ignored.


%
\begin{figure}[t!]
\begin{center}
 \includegraphics[width=30mm]{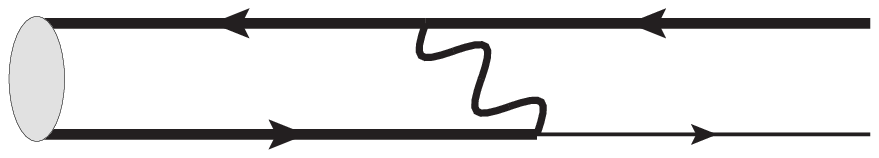}
 \hspace{5mm}
 \includegraphics[width=30mm]{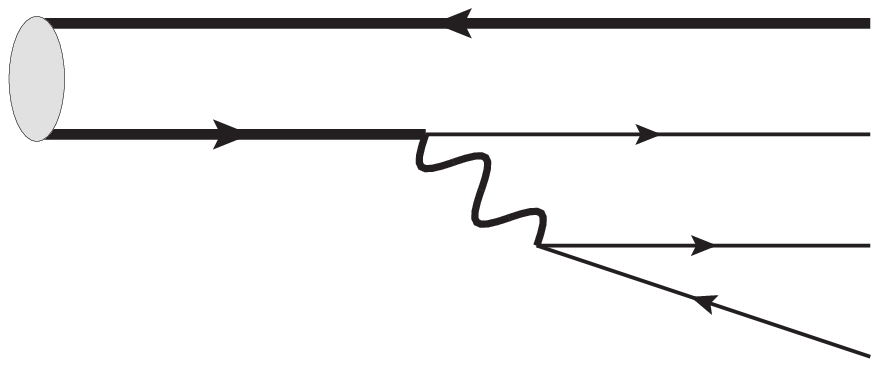}
 \caption{Weak decay modes:
  (a) exchange diagram with off-diagonal quark mixing element;
  (b) free quark decay.}
 \label{fig:weak}
\end{center}
\end{figure}

Another type of annihilation decay channel is
caused by weak boson exchange, where
an off-diagonal quark mixing element acts on
one of the bound quarks, as shown in Fig.~\ref{fig:weak}(a).
This is rather similar to the binding mechanism,
where the quark mixing elements are always within 4G.
Once the cross-generation interaction occurs,
the recoil due to the energy release from the
mass difference between the 4G quarks and lower generations
would eject the lighter quark and destroy the bound state,
followed by subsequent decay of the leftover 4G quark.
The lower generation quark mass is too light such that it
cannot bind with the heavy 4G quark by Yukawa coupling.

The decay width is calculated in terms of the decay constant in
Eq.~(\ref{eq:dec}) using the Fierz identity for the current products,
giving ($\hat m \equiv m/m_{\ome}$)
\begin{align}
 \Gamma(\ome\to t\bar{t'}) = \xi^2 |V_{tb'}^*V_{t'b'}|^2 \frac{G_F^2m_{\ome}^5}{192\pi}
 E(\hat m_{t},\hat m_{t'}),\\
 \Gamma(\ome\to b\bar{b'}) = \xi^2 |V_{t'b'}^*V_{t'b}|^2 \frac{G_F^2m_{\ome}^5}{192\pi}
 E(\hat m_{b},\hat m_{b'}),
\end{align}
where
\begin{align}
 E(x,x') &\; =
 \frac{\lambda(1,x^2,x^{\prime 2})}{2(1-2x^2-2x^{\prime 2} 
   )^2} \nn\\
\times & \Big[2 - 9 x^2 + 15 x^4 - 8 x^6 - 9 x^{\prime 2} + 18 x^2 x^{\prime 2} \nn\\
 &\; - 8 x^4 x^{\prime 2} - 16 x^6 x^{\prime 2} + 15 x^{\prime 4} - 8 x^2 x^{\prime 4} \nn\\
 &\; 
  + 32 x^4 x^{\prime 4} - 8 x^{\prime 6} 
  - 16 x^2 x^{\prime 6} 
 \Big].
\end{align}
with $\lambda(a,b,c)=\sqrt{a^2+b^2+c^2-2(ab+bc+ca)}$.
The charge conjugate decays have the same partial width as above.
We have actually performed a full calculation, but
set $M_W$, or weak coupling $g$, to zero at the end,
as we are concerned with longitudinal vector boson exchange.
We have also not distinguished between $2m_{t'}$ and $m_{\ome}$,
where the latter provides the scale parameter.
The decay rate depends on both the off-diagonal
quark mixing element $|V_{t'b}| = |V_{tb'}|$ as well as $\xi$.

In this mode, $\ome$ decays into on-shell $t\bar t'$ or $b\bar b'$ (and conjugate).
If we restrict to $t'\to bW$ and $b'\to tW\to bWW$ for the decay of 4G quarks,
the final state all end up as $bWbW$.
The signal is similar to $\ttb$ production, but the kinematical
distribution differs from the standard model counterpart.

\subsubsection{Free Quark Decay}\label{sub:fc}

A second type of decay mode is induced by the decay
of the constituent quarks, as illustrated in Fig.~\ref{fig:weak}(b).
It is similar to the weak boson exchange decay discussed
just before, and quantum mechanically speaking,
the exchanged boson escapes the system as an on-shell particle.
We call this the ``free" quark decay mode, even though
the decaying quark is bound.
The decay partial width in this mode depends crucially on $V_{t'b}$ and
$V_{tb'}$, but not strongly on the structure of $\ome$.
This last statement would no longer hold
when one enters the realm of deeply bound states, where
binding energy is much larger than the $-100$ GeV adopted here.

Inside the bound state system, the decay of the constituents is
suppressed by phase-space and time dilatation
effects~\cite{Sumino:1992ai,Jezabek:1992np}.
That is, the decaying quark constituent is off-shell.
However, for simplicity, we ignore these effects in
our rate calculation, and use
\begin{align}
 \Gamma_{\rm free} \simeq \frac{1}{2}\left[2\Gamma_{t'}+2\Gamma_{b'}\right]
 = \Gamma_{t'}+\Gamma_{b'},
\end{align}
where $\Gamma_{t'}$, $\Gamma_{b'}$ are given at Born level as
\begin{align}
 & \Gamma_{t'}(m_{t'}) = \left|V_{t'b}\right|^2 \frac{G_Fm_{t'}^3}{8\sqrt{2}\pi}
 F(\tilde m_W, \tilde m_b), \\
 & \Gamma_{b'}(m_{b'}) = \left|V_{tb'}\right|^2 \frac{G_Fm_{b'}^3}{8\sqrt{2}\pi}
 F(\tilde m_W, \tilde m_t),
\end{align}
with $\tilde m = m/m_{t'}$ or $m/m_{b'}$, and
\begin{align}
 F(x,y) = \left(1+x^2-2x^4-2y^2+x^2y^2+y^4\right) \lambda(1,x^2,y^2)
\end{align}
with $\lambda(a,b,c)$ as defined earlier.
Note that we have kept $M_W$ here, since the decay
process is quite similar to the familiar top quark decay.

The decay width of the 4G quarks is suppressed by
the small $V_{t'b}$ and $V_{tb'}$, but grows rapidly with
the 4G quark mass.
If $\ome$ decays through $t'$, the final state would be $bWbW$,
and $bWWbWW$ if decay is through $b'$.
The search of these signal can be along the standard 4G quark
search strategy~\cite{Chatrchyan:2011em},
except that, if heavy quark mass could be reconstructed,
then for example one $bW$ pair has a lower mass (due to binding energy)
than the other $bW$ pair from on-shell $t'$ decay~\cite{Sumino:2010bv}.

\begin{figure}[t!]
\begin{center}
 \includegraphics[width=35mm]{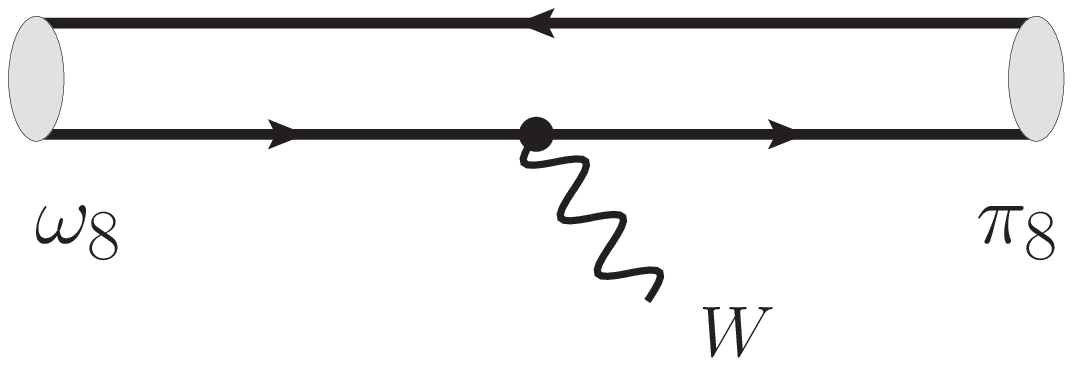}
 \hspace{3.5mm}
 \includegraphics[width=35mm]{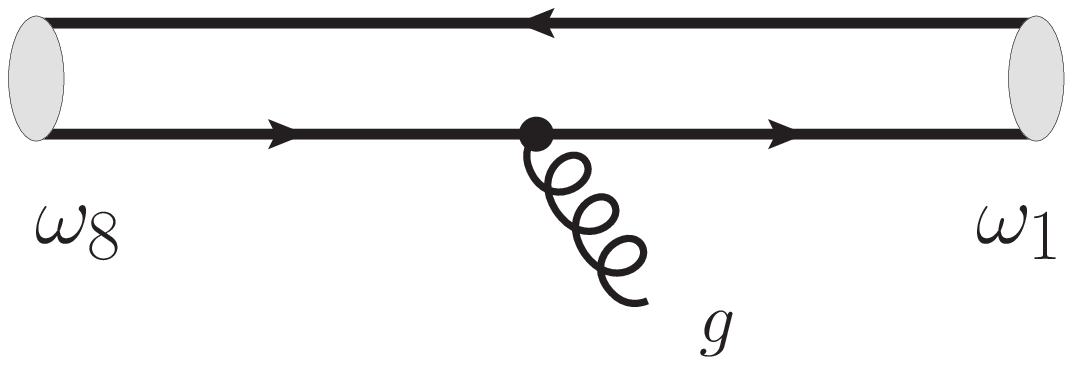}
 \caption{Meson transition currents for $\ome\to\pe$, and $\ome\to \omo$.
 }\label{fig:trans}
\end{center}
\end{figure}

\subsubsection{Meson Transition Decay}\label{sub:tran}

Finally, a third class of decay is for $\ome$ to turn into
other resonances. We consider the two channels of
$\ome\to\pe W$ and $\ome\to \omo g$. The other possible
channel $\ome\to \po g$ is forbidden by the heavy isospin.
The partial width for these decays depend on the
mass difference as well as the transition amplitude
of these resonances.
The meson transition to $\pe W$ would open only if
$m_{\ome}-m_{\pe}>m_W$.
For $m_{\ome}-m_{\pe}<m_W$, $\ome$ can decay into
$\pe \ell\nu_{\ell}$ or $\pe q\bar{q}'$ through the off-shell $W$ boson.
However, the partial width would be negligibly small.

We can write down a general vector to pseudoscalar transition
amplitude via (color singlet) vector or axial-vector currents as,
\begin{align}
 \langle \pe^a(p^\prime) |V^{\mu}|\ome^b(p)\rangle = &\;
  \frac{\delta^{ab}}{\sqrt{m_{\pe}m_{\ome}}}V(q^2)
 \,i\varepsilon_{\mu\nu\rho\sigma}\varepsilon^{\nu}
 p^{\rho}p^{\prime\sigma}, \\
 \langle \pe^a(p^\prime) |A^{\mu}|\ome^b(p)\rangle = &\;
  \frac{\delta^{ab}}{\sqrt{m_{\pe}m_{\ome}}}
 \Big[A_1(q^2) \,m_{\pe}m_{\ome}\varepsilon^{\mu} \nn\\
 + &\; A_2(q^2)\, p^\prime\cdot\varepsilon\, p^{\mu}
 + A_3(q^2)\, p^\prime\cdot\varepsilon\, p^{\prime\mu}
 \big],
\end{align}
where $p$, $p'$ are 4-momentum of $\ome$, $\pe$ respectively,
$\varepsilon$ is the polarization vector of $\ome$,
and $V$ and $A_{i=1,2,3}$ are form factors in
$q^2\equiv (p-p^\prime)^2$

A straightforward calculation gives ($\hat m \equiv m/m_{\ome}$)
\begin{align}
 \Gamma(\ome\to\pe W) = \frac{G_Fm_{\ome}^3}{32\sqrt{2}\pi}
 \, \frac{m_{\ome}}{m_{\pe}}
 \, W(\hat m_{\pe}, \hat m_{W}),
 \label{eq:opw}
\end{align}
with
\begin{align}
 W(x,y) = &\;
 |A_1|^2x^2(1-2x^2+x^4+10y^2-2x^2y^2+y^4)\lambda
  \nn\\
 &\; + {\rm Re}[A^*_1(A_2+A_3)]x(1-x^2+y^2)\lambda^3
  \nn\\
 &\; + \frac{1}{4}|A_2+A_3|^2\lambda^5 + 2|V^2| \, y^2\lambda^3,
 \label{eq:hpw}
\end{align}
where $\lambda=\lambda(1,x^2,y^2)$ as already defined.
Here, $M_W$ has to be kept, since transverse $W$ emission
has a $\beta$ phase space factor, while longitudinal $W$
emission has a $\beta^3$ factor and more suppressed.
Using the assumption that a free quark inside a meson
interacts with currents, the form factors are reduced to
$V = -1$, $2m_{\ome}m_{\pe}{A_1} = (m_{\ome}+m_{\pe})^2-q^2$,
${A_2} = -1$ and ${A_3} = 0$.
For simplicity, we use this limit in our numerical calculation,
and Eq.~(\ref{eq:hpw}) reduces to
\begin{align}
 W(x,y) \simeq &\; (1-2x^2+x^4+3y^2+2xy^2+3x^2y^2-4y^4) \nn\\
  &\; \times (1+2x+x^2-y^2)\,\lambda,
\end{align}
where $\lambda = \lambda(1,x^2,y^2)$.

\begin{figure*}[t!]
 \begin{center}
  \includegraphics[width=135mm]{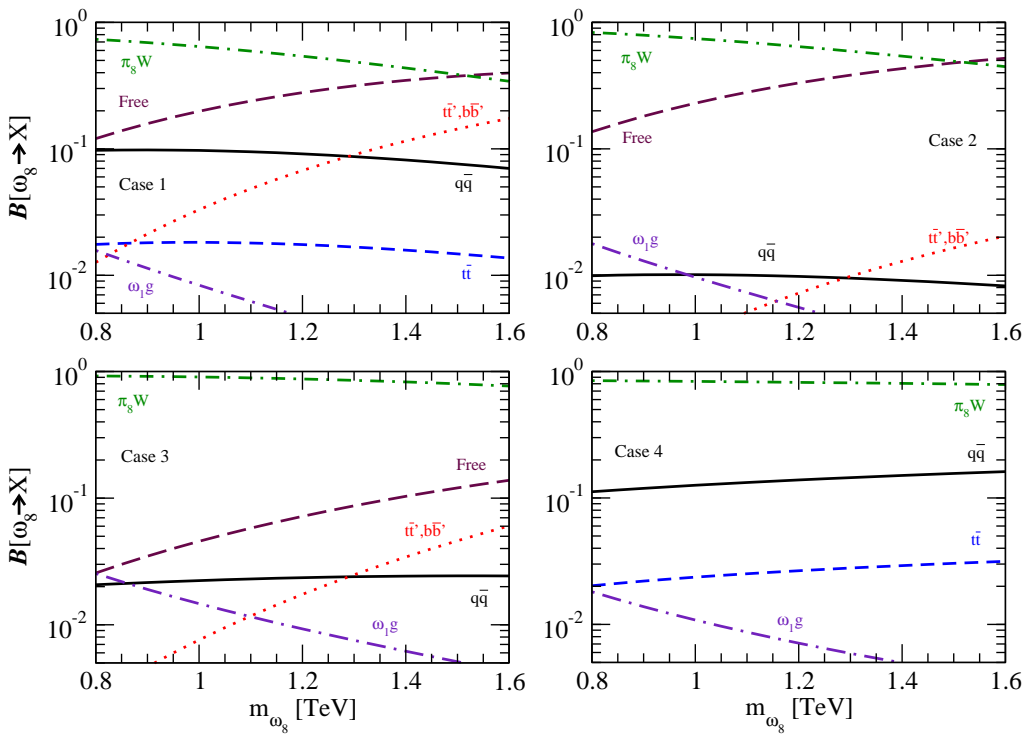}
  \caption{Branching ratio of $\ome$ as a function of
  $m_{\ome}$ for given parameter sets in Case~1 to 4.}\label{fig:br}
 \end{center}
\end{figure*}

A vector to vector transition amplitude via color octet vector current
is parameterized as
\begin{align}
 \langle \omo(p') |V^{\mu,a}|\ome^b(p)\rangle &\; =
 \frac{\delta^{ab}}{\sqrt6}
 \Big[
  \left({V_1}(q^2)\, p^{\mu} + {V_2}(q^2)\, p^{\prime\mu}\right)
  \varepsilon'\cdot\varepsilon \nn\\
 &\; + {V_3}(q^2)\, p\cdot\varepsilon'\, \varepsilon^{\mu}
   + {V_4}(q^2)\, p'\cdot\varepsilon\, \varepsilon^{\prime\mu}
 \Big],
\end{align}
where $\varepsilon$, $\varepsilon'$ are the
polarization vectors of $\ome$, $\omo$, respectively,
and ${V_{i=1\ldots 4}}$ are form factors in
$ q^2$. 
A straightforward calculation gives ($\hat m \equiv m/m_{\ome}$)
\begin{align}
 \Gamma(\ome\to\omo g) = \frac{\alpha_s}{18}\,
 \frac{m_{\ome}^2}{m_{\omo}} \,
 G(\hat m_{\omo}),
\end{align}
where we take the scale of the strong coupling constant
to be at the mass of $\ome$, and
\begin{align}
 G(x) = &\; \left(1-x^2\right)^3 \frac{|V_3|^2+x^2|V_4|^2}{2x^2}, \\
  \simeq &\; (1-x^2)^3/x.
 \label{eq:homg}
\end{align}
The second step follows from taking the free-quark limit
as described above, which reduces the form factors to
${V_1} = -{V_3} = -\sqrt{m_{\omo}/m_{\ome}}$
and ${V_2} = -{V_4} = -\sqrt{m_{\ome}/m_{\omo}}$.
We use this result, which is highly suppressed by phase space,
for our numerical estimation.

\subsubsection{Numerical estimates}

To perform numerical studies of the branching ratios and decay widths,
we finally have to specify the numerical values of the following parameters:
$\xi$, $V_{t'b}$ ($= -V_{tb'}$) and $\Delta{m}=m_{\ome}-m_{\pe}$ for
$\ome\rightarrow \pe W$ ($\Delta{m}=m_{\ome}-m_{\omo}$ for
$\ome\rightarrow \omo g$).
Without a full solution to the relativistic strong coupling
bound state problem, however, it is difficult to ascertain
the values for $\xi = f_{\ome}/m_{\ome}$ as well as the mass splittings,
and $m_{\ome}$ itself.
We therefore examine four sets of parameters as a survey,
\begin{itemize}
 \item Case~1~:\quad $\xi=0.1$,  $\Delta{m}=100$~GeV, $V_{t'b}=0.1$;
 \item Case~2~:\quad $\xi=0.03$, $\Delta{m}=100$~GeV, $V_{t'b}=0.1$;
 \item Case~3~:\quad $\xi=0.1$,  $\Delta{m}=200$~GeV, $V_{t'b}=0.1$;
 \item Case~4~:\quad $\xi=0.1$,  $\Delta{m}=100$~GeV, $V_{t'b}=0.01$.
\end{itemize}
These are chosen simply to emphasize large variety of possible dominant
decay channels.
In all the cases we set the binding energy of $\ome$ to
$m_{\ome}-2m_Q = -100$~GeV, which is much larger than QCD binding.
A different choice of the $\ome$ binding energy
changes our results only modestly.
For the choices specified in Case~1, we assume the larger decay
constant, ${\cal O}(100)$~GeV mass difference in the meson spectrum, and
$V_{t'b}$ is set to the nominal current upper limit~\cite{Holdom:2009rf}.
In Case~2, we examine the smaller decay constant.
Case~3 is for larger mass difference, and Case~4 is when $V_{t'b}$ is
more suppressed.
We will discuss the two different mass differences (vector--pseudoscalar
and octet--singlet) as variations.

We plot in Fig.~\ref{fig:br} the branching ratio of various $\ome$ decays
as a function of $m_{\ome}$ for Cases~1 to 4.
In Case~1, the dominant decay modes are the transition decay into $\pe W$,
especially for lighter mass region, and free quark decay,
i.e. via the decay of 4G constituent quark for heavier mass region.
The branching ratios of free quark decay and
the $V_{t'b}$-dependent annihilation ($W$ boson exchange) decay
increase with $m_{\ome}$, because $m_{\ome} \sim 2m_Q$
reflects a larger Yukawa coupling.
The $q\bar q$ is of order 10\% and drops slightly at higher $m_{\ome}$,
with $\ttb$ branching ratio a factor of 5 lower, at the percent level.
The transition decay into $\omo g$ is at the percent level or less.

In Case~2, because of the small decay constant, the annihilation decay
channels $t\bar t'$, $b\bar b'$, $q\bar q$ and $t\bar t$ are suppressed.
In this case, free quark decay and transition decay into $\pe W$ are
the two predominant modes.

In Case~3, the large mass differences enhance the
branching ratio of the transition decays, and
the $\pe W$ mode dominates.
The other transition decay into $\omo g$ can also be enhanced,
especially in the lighter mass region.
It could be that the mass difference of
only $m_{\ome}-m_{\omo}$ is large,
i.e. when the mass spectrum is like $m_{\ome}\simeq m_{\pe} > m_{\omo}$.
If so, $\pe W$ could be considerably suppressed,
and $\omo g$ would be more prominent,
especially for low $m_{\ome}$.

In Case~4, the free quark decay and the $V_{t'b}$ induced
annihilation decay are suppressed, due to small $V_{t'b}$.
The decay width of 4G quarks is also suppressed for the same reason.
In this case, the transition decay into $\pe W$ dominates,
and the annihilation decay into dijets can be sub-dominant
with branching ratio at ten percent order. However, this sensitively
depends on the $m_{\ome} - m_{\pe}$ mass difference,
as well as the decay constant.
If $\pe W$ becomes kinematically suppressed,
dijets would be dominant.

Let us summarize some general features regarding branching ratios.
Basically, the transition decay into $\pe W$ is large,
because of the large Yukawa coupling and
no suppression effect by bound state deformation.
This decay mode can be more enhanced if the mass difference is large,
but much suppressed if the mass difference is small, especially
if less than $M_W$, as we have seen.
Free quark decay has sizable contribution for the heavier mass region,
if $V_{t'b}$ is close to the current upper limit of 0.1.

\begin{figure}[t!]
 \begin{center}
\vskip0.13cm
  \includegraphics[width=75mm]{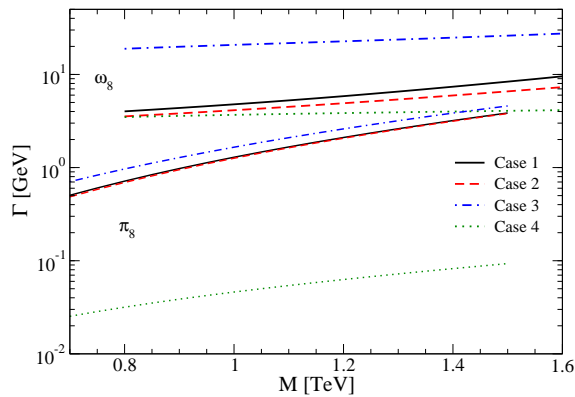}
  \caption{Total decay width of $\ome$ and $\pe$
   for the four parameter sets.
  The upper curves are for $\ome$, where Case 3 is
   enhanced by a larger $m_{\ome} - m_{\pe}$ mass difference.
  The lower curves, plotted 100 GeV less in mass, are for $\pe$,
   where Cases 1 and 2 are barely distinguishable.}
  \label{fig:wid}
 \end{center}
\end{figure}

In Fig.~\ref{fig:wid}, we show the total decay width of $\ome$
as a function of $m_{\ome}$ for the four parameter sets.
The decay width increases with $m_{\ome}$,
from a few GeV to around 10 GeV.
For Case 3,
due to the rapid decay into $\pe W$ from a
relatively large $m_{\ome} - m_{\pe}$ mass difference,
the width is at 20 GeV range,
and increases mildly with $m_{\ome}$.
Still, compared with its TeV scale mass,
$\ome$ is a heavy but narrow meson resonance.

We see that the binding energy, therefore strong
Yukawa dynamics, and flavor structure through $V_{t'b}$ all play crucial
role for the eventual phenomenology one expects at LHC.

\subsubsection{The Decay of $\pi_8$}

To be able to address LHC phenomenology, we need to treat $\omega_1$ and
$\pi_8$ further, as they may appear in $\omega_8$ decay final state.
From Fig.~\ref{fig:br} we see that in general $\pi_8W$ is the leading decay.
Note, however, that we have assumed $m_{\ome} - m_{\pi_8} = 100$ GeV.
The rate would drop sharply as this vector--scalar splitting
diminishes, and becomes practically negligible when $W$ turns virtual.
On the other hand, if the strong binding found by Ref.~\cite{Jain}
in the Bethe--Salpeter approach with $s$-channel subtracted has any bearing,
then $\pi_8$ may be deeper bound than $\omega_8$. For this situation, Case 3
stands as an illustration, where $\ome \to \pi_8W$ decay
would be preeminent.

In contrast, the process $\ome \to \omega_1g$ is never more than 10\%,
and more typically at $10^{-2}$ order or smaller.
The exception would be if $m_{\ome} - m_{\pi_8}$ is of order $M_W$ or less,
but $m_{\ome} - m_{\omo}$ is sizable
(plot of Case 3 in Fig.~\ref{fig:br}, but with $\ome \to \pi_8W$ removed).
Viewing this exception as unlikely, we relegate the discussion of
$\omega_1$ to a concurrent discussion of weak Drell--Yan production.

But we need to address how $\pi_8$ decays.
It is interesting that $\pi_8 \to \pi_1 g$ vanishes because
it is a $0^- \to 0^-$ transition, which we have verified by
direct computation.
The $W_L$ exchange diagram of Fig.~\ref{fig:weak}(a) is absent for
charged $\pi_8^\pm$ (i.e. $t'\bar b'$ and $b'\bar t'$ mesons) because of isospin,
while the $s$-channel annihilation is absent by the octet/isovector nature.
The upshot is that we are left with only two decay processes:
the familiar free quark decay, and a new type of decay,
$\pi_8 \to Wg$, where the $W$ is transverse.
The latter is an inverse process of $\ome \to \pi_8W$,
with $g$ replacing $\ome$. However,
the annihilation rather than transition nature implies
that $\pi_8 \to Wg$ is relatively suppressed.

\begin{figure}[t!]
 \begin{center}
  \includegraphics[width=72mm]{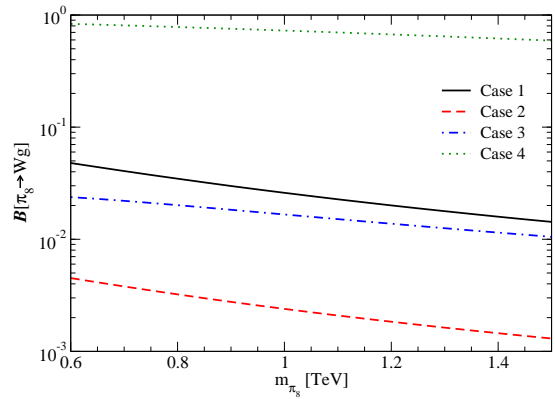}
  \caption{Branching ratio ${\cal B}(\pi_8 \to Wg)$
   for the four parameter sets, ranging from
   predominance (Case 4), to below $10^{-2}$ (Case 2).}
  \label{fig:pi8}
 \end{center}
\end{figure}

Direct computation gives
\begin{align}
 & \Gamma(\pi_8 \to Wg) \nonumber \\
 =\ & 2 \xi_\pi^2 \alpha_s\alpha_W\, m_{\pi_8}
 \frac{(1 - M_W^2/m_{\pi_8}^2)^3}{(1 + 4m_Q^2/m_{\pi_8}^2 - 2M_W^2/m_{\pi_8}^2)^2},
\end{align}
where $\xi_\pi = f_{\pi_8}/m_{\pi_8}$ is the $\pi_8$ decay constant
normalized by $\pi_8$ mass.
The rate of $\pi_8\to Wg$ is suppressed by $\alpha_W/\alpha_S\, n_f$
compared to the rate of $\ome \to q\bar q$ of Eq.~(5).
We have checked explicitly that longitudinal $W$ emission again vanishes.

We plot the $\pi_8$ width in Fig.~\ref{fig:wid}, and
the ${\cal B}(\pi_8\to Wg)$ branching fraction, in Fig.~\ref{fig:pi8}.
The width is at GeV order, narrower than $\omega_8$,
but could be much smaller if a small $V_{t'b}$ suppresses the
free quark decay widths.
Thus, $\pi_8\to Wg$ decay branching fraction is below 10\%,
and much smaller for Case 2 (suppressed by a smaller decay constant).
For Case 4, the small $V_{t'b}$ case,
$\pi_8\to Wg$ could dominate.

\begin{figure}[t!]
 \begin{center}
  \includegraphics[width=70mm]{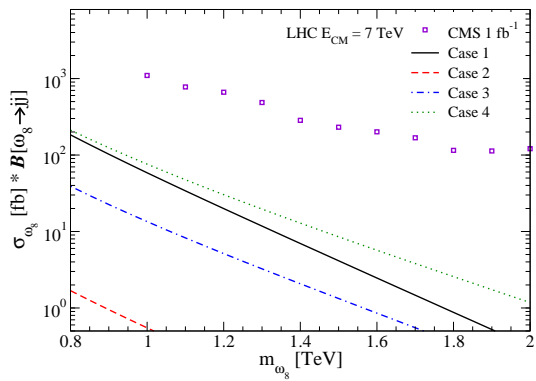}
  \caption{Production cross section times branching ratio
  for $\ome$ into dijets, for the four parameter sets at the LHC 7~TeV.
  CMS upper limit~\cite{Chatrchyan:2011ns} for the cross section of
  dijet resonance production is also plotted.}\label{fig:jj}
 \end{center}
\end{figure}

\section{\label{sec:Discussion} Discussion\protect\\}

We now discuss the possible phenomenology,
as well as other issues.

Let us first comment on the dijet decay of $\ome$,
which could appear as a dijet resonance.
Absence of dijet features in Tevatron and LHC searches
constrains or rules out any model with particles in TeV range.
In Fig.~\ref{fig:jj}, we plot the total cross section times
dijet branching ratio for $\ome$ production at the LHC 7 running at TeV,
as a function of $m_{\ome}$ for the four parameter sets.
We compare with the CMS dijet resonance search~\cite{Chatrchyan:2011ns}
with ${\cal L}=1$~fb$^{-1}$ data at $\sqrt{s}=7$~TeV.
We note that the data includes acceptance cuts for dijets,
but our model numbers do not, which makes the comparison conservative.
We find the cross section times dijet branching ratio are
at least an order of magnitude lower than the current upper limit
for all four Cases, with Case~4 the largest.
Even with $\ome \to \pe W$ channel removed
making $\ome \to q\bar q$ the leading decay,
one is still below the CMS limit.
Let us call this special situation Case~$4'$.
Our results show that, while a narrow resonance signal
might start to show up with a considerably larger data set,
it could show up soon for Case~$4'$. That is,
if $\ome \to \pe W$ decay is forbidden,
while free quark and exchange decays are suppressed by small $V_{t'b}$.
The need for these two conditions to be met, however,
makes this possibility not particularly likely.

In general, the $\ome \to \pe W$ decay is the dominant $\ome$ decay mode,
unless it is kinematically suppressed by $m_{\ome} - m_{\pe}$
being too close to, or smaller than, $M_W$.
We have investigated the decay of $\pi_8$ itself,
and found that it is dominated by free quark decay,
with $\pe \to Wg$ subdominant for Cases 1--3.
It is important now to spell out the isospin nature:
isosinglet $\ome^0 \to \pe^\pm W_s^\mp$, $\pe^0 W_s^0$ with a $2 : 1$ ratio,
where the subscript $s$ indicates a relatively soft vector boson,
and $W^0$ stands for the $Z$ boson, as we
have ignored heavy isospin breaking.
Thus, the signature for $q\bar q \to \ome \to \pe W$ leads to
$bW^+\bar b' W_s^-$, $tW^-\bar t' W_s^+$, and
$\{bW^+\bar t',\; tW^-\bar b'\}W_s^0$, plus c.c.
Except for the last case with presence of a $Z$ boson
(the identification via dileptons would be costly in branching fraction),
the additional $W^\pm$ which is relatively soft
makes it an even more complicated signature than
direct open $Q\bar Q$ production.
The same sign dilepton approach~\cite{Chatrchyan:2011em}
remains a good one, but more jets would be present.
Note that the $Z$ is usually vetoed against in same sign dilepton studies.

However, if boosted $W$ and top jets can be exploited
to isolate the $bW^+\bar b'$ and $tW^-\bar t'$
(both in $bWtW$ final state),
then the associated soft $W_s^\mp$ could be
an extra tag for $\pi_8^\pm W^\mp$ production.
Besides the relatively low $p_T$ $W$,
the 7-jet system has rich imbeddings of $W$, $b$ and $t$ jets.
It can be part of the $t\bar t W$ search program,
which would in any case be a background.
But the multi-jet system is of rather high mass,
and with different signature content, that one
may be able to separate.
If the total jet mass resolution is good,
one may discover both the $\pi_8^\pm$ in
multijets with a soft $W$ tag, with the
$\pi_8^\pm$ and the $W^\mp$ reconstructing
to $\omega_8$.

The LHC experiments should also look for a $Z$ boson
associated with high jet multiplicity,
perhaps with a hint of unusual $ZW$ plus multijets
as backdrop.
Given that $\pe$ is very narrow, one could look
for a relatively soft $Z$ recoiling from
6 or 8 hadronic jets with very high,
but relatively specific mass,
containing substructures of multiple boosted $W$ jets
or boosted top jets. The whole event, $Z\, +$ 6 or 8 jets,
would also reconstruct to a narrow resonant mass.

One might think that $\gamma\, +$ multijets
can be similarly pursued. However, Eq.~(\ref{eq:opw})
does not apply to $\ome \to \pe \gamma$:
the $V(q^2)$ term vanishes with $M_W \to 0$.
The photonic decay involves heavy quark spin flip,
hence suppressed by $m_Q^2$.
Treating $m_{\ome} + m_{\pe} \simeq 2m_{\ome}$, since we
allow $\Delta m = m_{\ome} - m_{\pe} \sim 200$ GeV at best,
the radiative rate is $\Gamma(\ome \to \pe \gamma)
 \simeq \frac{\alpha}{3}\frac{(\Delta m)^3}{m_Q^2}$.
Since a larger $\Delta m$ (Case~3) gives a larger
total $\ome$ width, we find that ${\cal B}(\ome \to \pe \gamma)$
is always below the percent level.
However, photon detection does not suffer from
the factor of 0.06 as for $Z\to \ell^+\ell^-$ detection.
Hence, the LHC experiments might also consider
$\gamma\, +$ multijet studies.

Case 4 offers yet again an intriguing signature,
assuming $\ome \to \pe W$ decay dominance.
From Fig.~\ref{fig:pi8} we see that $\pe \to Wg$ is dominant,
as free quark decay is suppressed by a small $V_{t'b}$.
One therefore has a unique signature of $W_sWg$
 ($W_s^\pm W^\mp g$ and $Z_s Zg$).
Here, one $W$ is soft, and the other hard,
with $p_T$ greater than 500 GeV, accompanied by
a similarly hard gluon. Both $W$'s tend to be transverse,
but the hard $W$-jet and the gluon jet would
form a rather narrow resonance!
This case offers dramatic signature
and should be quickly searched for.
The production cross section, of course,
is modulated by $f_{\ome}$ (see Fig.~\ref{fig:pro}).
Case 4, which is the limiting case of small $V_{t'b}$,
has better likelihood than the even more special
Case 4$^\prime$ discussed earlier.

Leaving Case 4, i.e. if $V_{t'b}$ is closer to 0.1,
the next prominent decay compared with $\ome \to \pe W$
in general is ``free" quark decay,
i.e. one of the bound $t'$ or $b'$ quarks decays,
dissolving the bound state system.
This could be practically the only other decay mode,
if the exchange and annihilation decays are
suppressed by a small decay constant (Case 2).
The signature is $q\bar q \to \ome \to bW\bar t'$,
$tW\bar b'$ (plus c.c.), where the notation implies the
associated $\bar t'$ or $\bar b'$ decays on-shell,
but the $bW$ and $tW$ are decay products of
a bound, somewhat off-shell $t'$ or $b'$ quark,
which is not too different from open $Q\bar Q$ production.
Unless the on-shell nature is used in the direct
search, the search limits would not be affected.
However, once 4G quarks are discovered, one
should check whether, for some fraction of the events
(depending on resonance vs open $Q\bar Q$ production ratio),
one of the heavy quarks is in fact off-shell,
which would be an indication for bound state phenomena.
We remark that open $Q\bar Q$ production at the LHC
is dominantly through $gg$ fusion, while the $\ome$
resonance production is through $q\bar q$ fusion.
There is little resonance phenomena in $gg \to Q\bar Q$
via Yukawa effects. In fact, in the $\eta_{(8)}$ channels,
it is even repulsive. Thus, even above the unitarity bound,
standard search can continue, except that there may be
some ``anomalies" as we have discussed, if any thing is
found at all.

We have already dealt with the special case of
dijet resonance for Case 4. Dijets from
$q\bar q \to \ome \to q\bar q$ tend to be
subdominant in all other Cases (i.e. 1--3),
but it could be at 10\% level.
It would provide a spectacular dijet resonance signal, and
definite measurement~\cite{Kats:2009bv} of resonance mass,
and spin if there is good signal over background.
If the branching ratio could be measured in some way,
one could access the important decay constant.
In general, a resonance would also appear in $t\bar t$
(boosted top jets), with cross section 1/5 the dijet resonance.

Finally, there is also the exchange decay
to $t\bar t'$ and $b\bar b'$, which is
a subdominant channel typically below 10\%.
It mimics ``single $t'$ ($b'$)" production,
and can be studied that way.
But an associated boosted top,
or a high $p_T$ $b$-jet that tags a resonant $tW$,
could be quite distinct.

We offer some remarks on the 
$A_{\rm FB}^{t\bar t}$~\cite{AFBtt} and $Wjj$~\cite{Wjj} 
anomalies at the Tevatron.
Naively, one might think that the presence of
resonance production of $t\bar t$ could be
relevant for $A_{\rm FB}^{t\bar t}$.
But $\ome$ has same quantum numbers as the gluon,
i.e. the coupling to $t\bar t$ is fully vector.
Thus, it cannot generate $A_{\rm FB}^{t\bar t}$
suggested by Tevatron data.
For the $Wjj$ anomaly,
the Yukawa bound resonances are so massive,
they can have nothing to do with it.

This brings us to comparing with Technicolor (TC) models.
Low scale TC has been invoked~\cite{Eichten:2011sh}
for the $Wjj$ anomaly suggested by CDF.
Our $\ome$ and $\pe$ are Yukawa bound $Q\bar Q$ mesons
with an operative heavy isospin, from degenerate chiral quark
doublet $Q$ not too far above the unitarity bound.
Thus, our $\ome$ and $\pe$ mesons are much heavier than those
in low scale TC.
For more generic TC models~\cite{Martin:2008cd},
since ``walking" is generally required,
the technipion $\pi_T$ tends to be closer to the technirho $\rho_T$
in mass such that $\rho_T \to \pi_T\pi_T$ is absent, while
(near) degeneracy of $\omega_T$ and $a_T$ with $\rho_T$ is
also often invoked. The signature for these technimesons
are typically $WZ$, $W\gamma$ and $Z\gamma$.
Thus, not only the spectrum is rather different ---
absence of $\rho$ and $a$ mesons --- the decay signature
is also in strong contrast.
The bound states due to strong Yukawa coupling, which follow
simply from the existence of new heavy chiral quarks
without assuming new dynamics, should be
easily distinguished from Technicolor.

The Yukawa-bound ultraheavy mesons are also quite distinct from
QCD bound states. Not only there is the absence of $\eta$
(where $gg$ fusion would be possible) and $\rho$ type mesons,
they have a much larger binding energy, and are much
smaller in size. This is brought about by not only
a strong coupling constant, but facilitated by a $\gamma_5$
coupling due to Goldstone or longitudinal vector bosons;
the $0^+$ Higgs boson, being heavy, would in fact be subdominant.
Thus, the tight bound states involve ultrarelativistic
motion of its very heavy constituents, hence somewhat
counterintuitive.
By far we have not attempted any actual solution of
the bound state problem here. We therefore chose to
remain close to the unitarity bound, considering
bound state masses not more than 1.5 TeV.
We have chose to parameterize with a few key parameters.
Our numerics, and associated phenomenology, should
be viewed as only illustrative,
with the \emph{key parameters to be determined by experiment}.

What mass scale would Nature actually choose, if she
so chooses to offer an extra chiral doublet above the existing
three generations? It may be related to the electroweak
symmetry breaking through $\bar QQ$ condensation.
It could in principle lead to very deeply bound states,
with binding energy approaching $m_Q$ order or more.
We have only scratched the surface of
Yukawa-bound heavy mesons, the treatment of which would
require genuine nonperturbative methods,
such as~\cite{dlin} lattice field theory.
Paradoxically, it is not impossible that heavier quark masses
than considered here could result in lower heavy meson masses.
Again, experiment might take the lead here.

\section{\label{sec:Conclusion} Conclusion and Outlook\protect\\}

With the experimental limits on sequential
chiral 4th generation already at 500 GeV,
i.e. at the doorsteps of the unitarity bound,
we have considered the possibility of new $\bar QQ$ mesons
bound by strong Yukawa couplings.
Comparing a relativistic expansion approach
(which indicated nonapplicability),
versus a relativistic Bethe--Salpeter equation approach,
we chose to illustrate what might appear in early data of
LHC running, i.e. bound states just above the TeV scale,
but with relatively complicated decay final states.

Electroweak precision tests suggest a new (heavy)
isospin symmetry, such that the leading production
would be a color octet, isosinglet vector meson, which
we call $\ome$. It can be produced via $q\bar q \to \ome$ fusion,
through an unknown decay constant, $f_{\ome}$.
For decay, the key parameters besides $f_{\ome}$ are
the quark mixing element $V_{t'b}$, and the
mass differences $m_{\ome} - m_{\pe}$ and $m_{\ome} - m_{\omo}$
where $\pe$ is a heavy color octet ``pion" and
$\omo$ a color singlet ``omega".
We find the leading decay is likely $\ome \to \pe W$, while
the other transition channel into $\omo g$ is relatively suppressed.
The other leading decay is ``free", or constituent, quark decay.
Illustrating with four Cases for large/small decay constant,
nominal/suppressed mass differences or $V_{t'b}$,
together with the two decay channels of free quark decay and
$Wg$ decay of $\pe$, we considered the possible LHC phenomenology.
We find in general the $\omega_8$ to be narrow
compared to its mass.

The special case of small $V_{t'b}$ leads to
two possible spectacular signatures.
One is $q\bar q \to \ome \to \pe W_s \to W_s Wg$, where
a massive back-to-back $Wg$ system is accompanied by a
relatively soft $W$.
This mode could become suppressed if $m_{\ome} - m_{\pe}$
is close to or less than $M_W$. Then,
one could have a dijet resonance close to current LHC limits,
and a narrow dijet resonance could appear soon.
In the general case, dijet signal is suppressed by a
decay branching ratio.

Other than the above two (perhaps unlikely) spectacular signals,
the generic leading decay is $\ome \to \pe^\pm W_s^\mp$, $\pe^0 Z_s^0$,
followed by free quark decay of $\pe$
($\pe \to Wg$ is typically subdominant).
This leads to possible multijet signals with an associated
relatively soft $W$ or $Z$ tag, where the multijet system
is very massive, and with multiple $b$, $W$ and $t$ jet substructures.
If such massive multijet systems can be studied,
one could possibly reconstruct both the $\pe$ and $\ome$ resonances.
Assuming single channel dominance,
one can measure the meson decay constant
by the total cross section.

If  $\ome \to \pe W$ is suppressed by kinematics, however,
the likely leading decay would be by the constituent heavy quark decay,
which is very similar to standard $Q\bar Q$ signal,
except one heavy quark decays somewhat off-shell.
Since in any case the leading $gg \to Q\bar Q$ fusion
does not exhibit resonance phenomena,
\emph{the current 4th generation $t'\bar t'$ and $b'\bar b'$
search can continue beyond the unitarity bound}.
But if 4G quarks are discovered, then some good fraction of
the events would have one quark decaying below threshold,
indicating bound state phenomena.
One, of course, would have to disentangle also $\ome \to \pe W$,
as already discussed.

We have provide some definite signatures for Yukawa-bound
heavy $Q\bar Q$ mesons in the 1 to 1.5 TeV range.
But our study is only of precursory nature.
As LHC energy increases, and with higher luminosity,
it could uncover new chiral quarks above the unitarity bound,
with new unusual bound states.
One could probe into the truly nonperturbative regime,
which our results only offer a glimpse of what might happen.
There may be a host of new heavy mesons awaiting us
beyond the horizon.

\vskip0.3cm
\noindent{\bf Acknowledgement}.
We thank K.-F. Chen, A. Djouadi, K. Ishiwata, B. Kniel, T. Kugo,
M.B. Wise, B.-L. Young and C.-P. Yuan for discussions,
and K. Hagiwara for encouraging comments.
WSH thanks the National Science Council for Academic Summit grant
NSC 99-2745-M-002-002-ASP, and
TE and HY are supported under NSC 100-2811-M-002-061
and NSC 100-2119-M-002-001.


\end{document}